\documentclass[12pt]{iopart}

\usepackage{graphicx}

\begin{document}
\newcommand{\shorttitle}{Scalable ion traps} 
\title[\shorttitle]{Scalable ion traps for quantum information processing}

\author{
J. M. Amini\footnote{Current address: Georgia Technology Research Institute, Signature Technology Laboratory, CRB Building, 400 Tenth Street, Atlanta, Georgia 30318.}
H. Uys,
J. H. Wesenberg\footnote{Current address: Department of Materials, University of Oxford, OX1 3PH, United Kingdom.},
S. Seidelin\footnote{Current address: Institut Neel-CNRS, BP 166, 25, rue des Martyrs, 38042 Grenoble Cedex 9, France.},
J. Britton\footnote{Current address: National Institute of Standards and Technology, Quantum Electrical Metrology Division, 325 Broadway, Boulder CO 80305.},
J. J. Bollinger,
D. Leibfried,
C. Ospelkaus,
A. P. VanDevender,
and
D. J. Wineland}
\address{
National Institute of Standards and Technology, 
Time and Frequency Division
325 Broadway, Boulder CO 80305
}

\begin{abstract}
We report on the design, fabrication, and preliminary testing of a 150 zone array built in a `surface-electrode' geometry microfabricated on a single substrate.  We demonstrate transport of atomic ions between legs of a `Y'-type junction and measure the in-situ heating rates for the ions. The trap design demonstrates use of a basic component design library that can be quickly assembled to form structures optimized for a particular experiment.
\end{abstract}


\tableofcontents

\renewcommand{\leftmark}{\shorttitle} 

\section{Introduction}

The basic components for a quantum information processor using trapped ions have been demonstrated in a number of experiments \cite{Blatt08.Nature.453.1008, Monroe08.PhysicsWorld.Aug, Haeffner08.PhysicsReports.469.155}. To perform complex algorithms that are not tractable with classical computers, these components need to be integrated and scaled to larger numbers of quantum bits (qubits).  Both integration and scaling can be achieved by making trap arrays with many zones.  In one possible scheme, information is shared between zones by physically transporting the ions between trapping zones that have various specialized functions, such as detection, storage, and logic gates \cite{Kielpinksi02.Nature.417.709, Kim05.QIC.5.515, Steane07.QIC.7.171}.  We report here on the design, fabrication, and preliminary testing of a large array built in a `surface-electrode' geometry \cite{Chiaverini05.QIC.5.419,Seidelin06.PRL.96.253003} and report on the first transport of atomic ions through a surface-electrode trap junction.  The trap is composed of 150 zones and six `Y' type junctions and is in principle scalable to an arbitrarily large number of zones.  It demonstrates use of a basic component design library that can be quickly assembled to form structures designed for a particular experiment or, in the future, a particular algorithm.  Microfabricated on a single substrate, the traps are amenable to rapid mass fabrication.

Ion trap design, fabrication, and characterization for quantum information experiments can be a difficult and time-consuming process.  With a design library and fabrication techniques as demonstrated in this paper, we illustrate the use of pre-designed and pre-characterized, modular components that can be assembled into trap designs for specific experiments. The design library includes components for Y junctions, loading, transport, and generic `experiment' regions.  We combined six of the Y junctions into an hexagonal ring that includes two loading components and two experiment regions. Fabrication of five of the unmounted traps on a single 76 mm diameter wafer took one week. 

\begin{figure*}
  \hspace{1in}\includegraphics{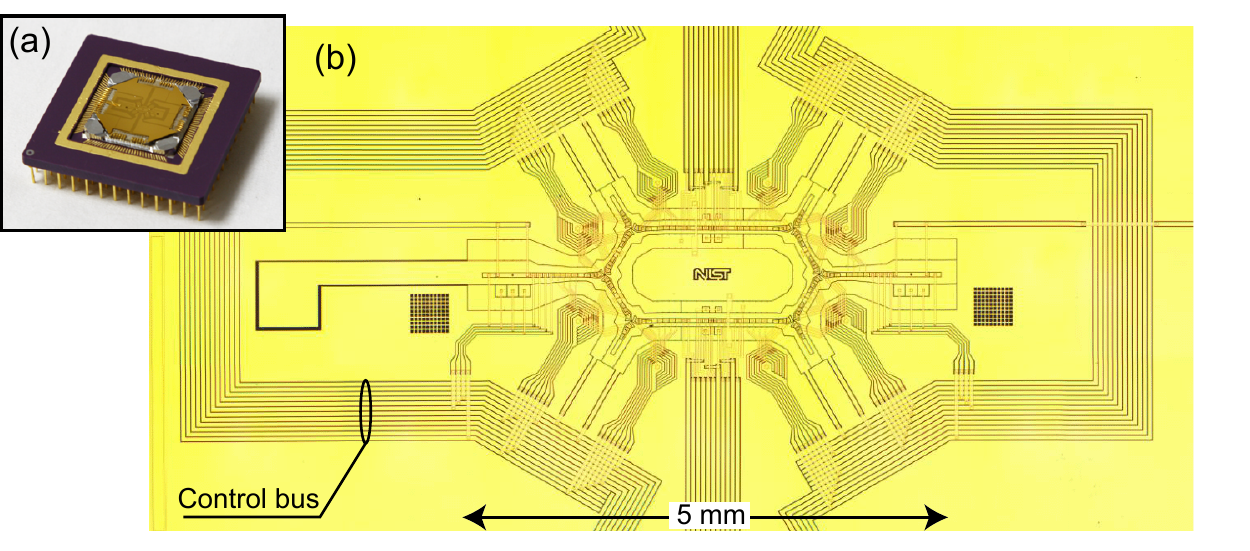}\newline
  \hspace*{1in}\includegraphics{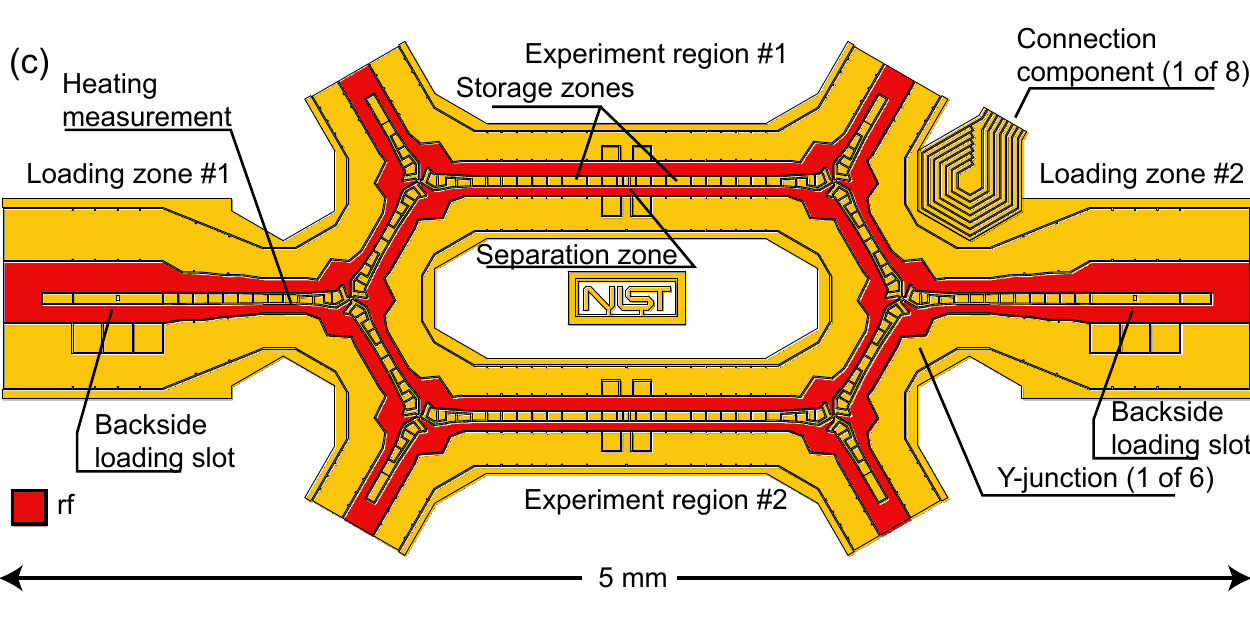}
  \caption{Photographs of (a) the trap mounted in a 120 pin PGA (pin grid array) carrier and (b) the active region of the fabricated trap. The design of the trap electrodes is shown in (c). This design features 150 transport/storage/probing zones, six Y junctions, and separation regions in the top and bottom paths of the hexagon. The trapping regions and the control bus are connected using standardized connection components that sit in crooks of the Y junctions.  Future designs could scale the structure by repeating the hexagonal pattern or could create new patterns using the same component library.}
  \label{image}
\end{figure*}

\begin{figure*}
  \hspace{1in}\includegraphics{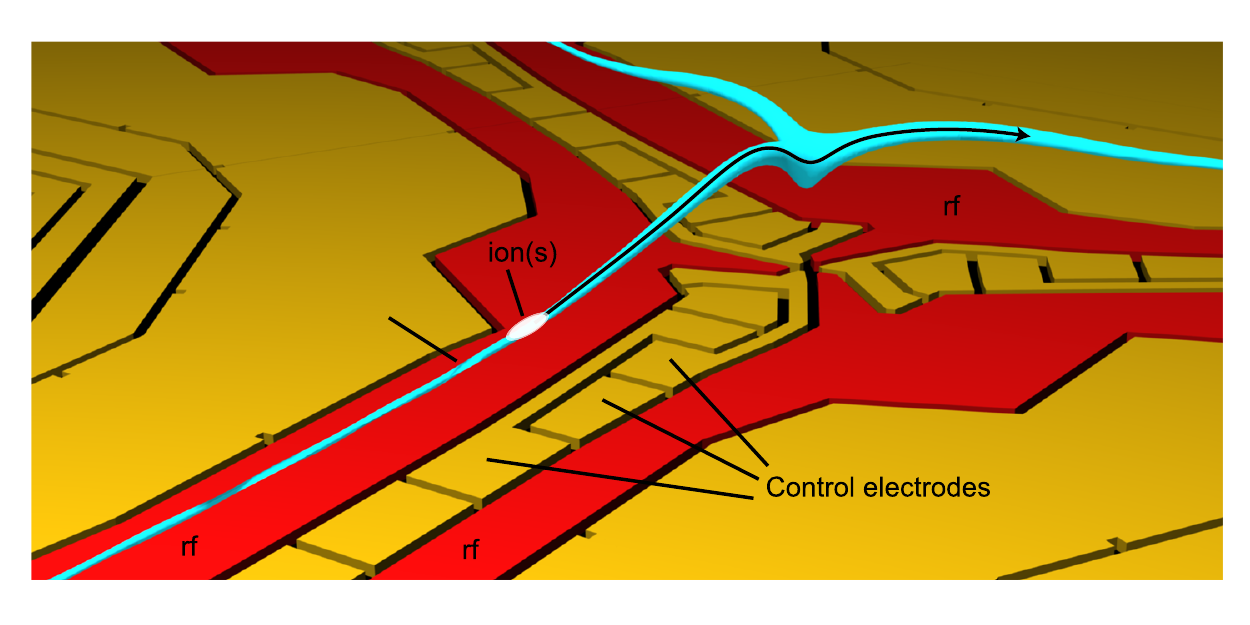}
  \caption{Example rf ponderomotive potential contour. For the experimental conditions used when testing the trap (113 V peak rf at 90.7 MHz and trapping $^{24}$Mg$^+$), this contour corresponds to 1.3 meV.  By changing the potentials on the control electrodes, the ion is moved along the two-dimensional confining tube formed by ponderomotive potential.  Junctions join multiple tubes and the control potentials move the ion between the junction's legs.}
  \label{potentials}
\end{figure*}

The ion trap described in this paper and shown in figure~\ref{image}(a) and (b) is a planar version (surface-electrode trap) of a radio-frequency (rf) Paul trap \cite{Chiaverini05.QIC.5.419,Seidelin06.PRL.96.253003}  that confines ions by combining a ponderomotive potential generated by an rf electric field  and static electric potentials, as shown in figure~\ref{potentials}.  In the configuration shown in figure \ref{image}, the ponderomotive potential (also called the pseudopotential) does not form a fully three-dimensional trapping potential.  Instead, it forms a confining tube and control potentials applied to segmented electrodes confine ions along the axis of the tube(s). By changing the control potentials slowly with respect to the rf period, the ions can be transported along the tube.  For added versatility, the trap includes junctions that combine multiple pseudopotential tubes and allow the ions to switch between multiple paths.  

\section{Fabrication}

\begin{figure}
  \hspace{1in}\includegraphics{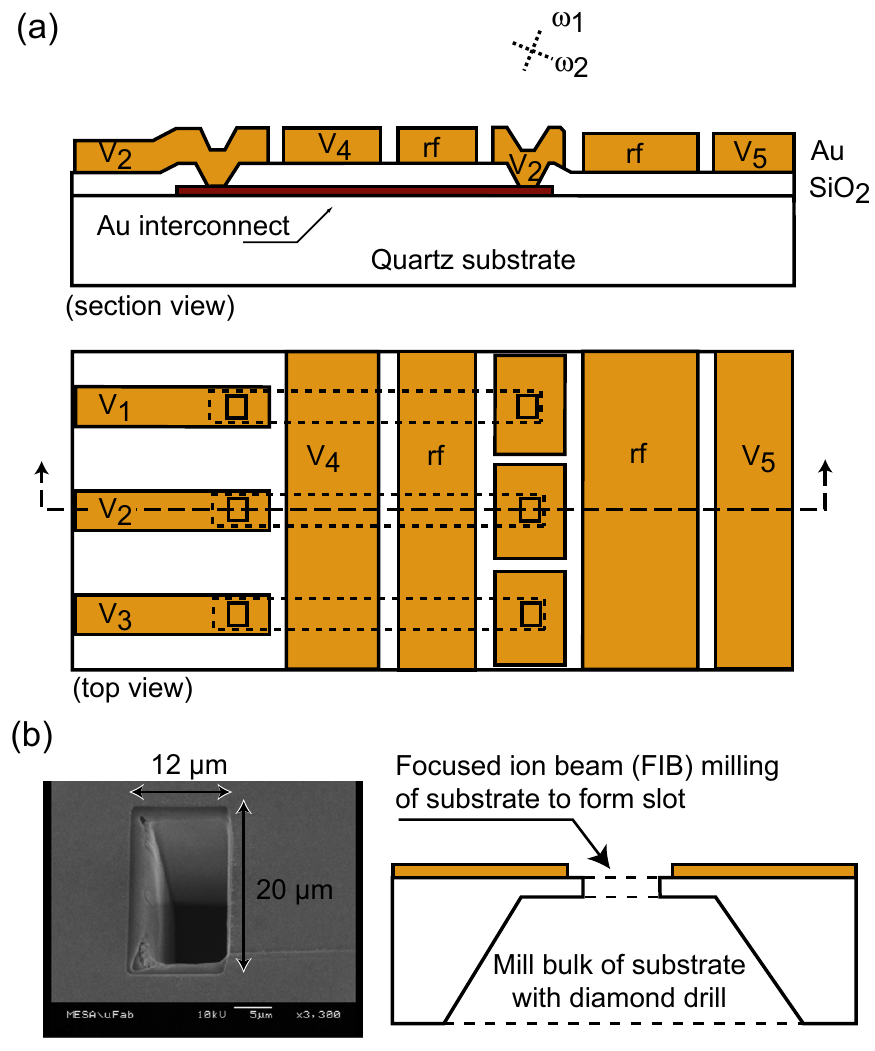}
  \caption{(a) Trap fabrication on a quartz substrate.  The lower metal layer is evaporated gold (with titanium adhesion layers on both sides).  The upper metal layer is evaporated gold with a titanium adhesion layer. The two metal layers are separated by a chemical-vapor-deposited silicon dioxide layer.  The oxide layer is plasma etched to form vias between the layers. The rotated `X' in the section view indicates the axes of the rf quadrupole electric field with the origin at the rf field null. (b) Backside loading slot. Neutral $^{24}$Mg enters the trapping region through a slot in the quartz wafer and is then photoionized.}
  \label{ChipSection}
\end{figure}

The trap fabrication is based on gold-on-quartz structures reported on in \cite{Chiaverini05.QIC.5.419,Seidelin06.PRL.96.253003} where a quartz wafer (which has low rf loss) is coated with a patterned gold conducting layer. The monolithic construction of surface-electrode traps provides a basis for scalable ion trap structures.  However, as the trap complexity increases, distributing the potentials to the control electrodes becomes increasingly difficult.  The design of the trap described here could not be realized using a single conducting layer because control electrodes between the rf `rails' form isolated islands surrounded by other control electrodes.  To counter this problem, we extended the single conducting layer fabrication used in \cite{Seidelin06.PRL.96.253003} to multiple conducting layers, as shown in figure~\ref{ChipSection}(a).  The top conducting layer forms a nearly continuous conducting plane that shields the ion from the potentials on the second (lower) conducting layer (see \cite{Leibrandt09}). 

The fabrication process begins with an 380 $\mu$m thick amorphous quartz wafer.  The first conducting layer is 300 nm of evaporated Au with 20 nm Ti adhesion layers deposited on both sides of the Au.   These layers are lithographically patterned and etched (wet etch for the Au and plasma etched for the Ti) to form the interconnects (figure~\ref{ChipSection}).  A 1 $\mu$m layer of SiO$_2$ is then deposited by chemical-vapor-deposition, forming the insulation between the two layers. The Ti layer deposited on top of the Au acts as an adhesion layer for this oxide. The vias in the oxide are plasma etched with a process that results in sloped side walls, ensuring electrical continuity to the conducting layer. The top conducting layer is 1 $\mu$m of evaporated Au with a 20 nm Ti adhesion layer, patterned using the liftoff technique \cite{Jaeger02.IntroMicroFab}.

The trap was loaded by passing a neutral flux of $^{24}$Mg through slots in the wafer. To form these slots, tapered channels through the quartz wafer were mechanically drilled from the backside of the wafer to within 30 $\mu$m from the top front surface at locations under the two loading zones.  The remaining 30 $\mu$m membrane was then milled using a focused ion beam to form a 12 $\mu$m by 20 $\mu$m  slot in the surface connecting to the drilled channel.

\section{Trap geometry}

The trap design, shown in figure~\ref{image}(c), incorporates use of a library of patterns that can be connected together to form more complex structures. The core of the trap design are the six Y junctions that are assembled to form a hexagonal ring. Two loading regions at either end feed ions into the hexagonal ring.  Inserted into two legs of the hexagon are components that can combine and separate pairs of ions for entangling and distributing the ions. Except for the loading regions, the outward legs of the hexagon are terminated in this design. However, these legs could be extended in future designs to integrate more hexagonal rings or other components such as memory storage regions.  

\begin{figure}
  \hspace{1in}\includegraphics{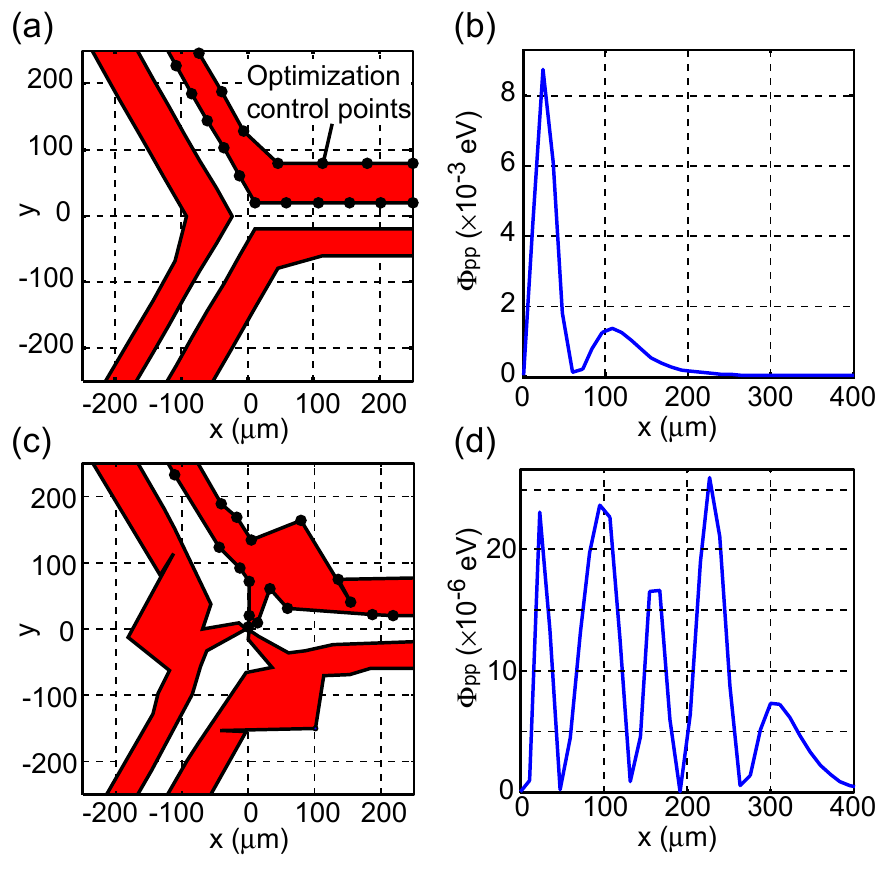}
  \caption{Optimization of junction rf electrodes (shown in red). (a) Initial shape used for optimization and (b) the ponderomotive potential $\Phi_{pp}$ at the center of the ponderomotive tube.  The horizontal axis of (b) is the position along the ponderomotive tube starting at the junction center ($x=0$) and proceeding outward along the tube to the right. The ponderomotive potential is evaluated for the experimental conditions (113 V peak rf at 90.7 MHz and trapping $^{24}$Mg$^+$). (c) Optimized junction shape and (d) its ponderomotive potential that is 300 times smaller than the unoptimized case shown in (b).}
  \label{Optimization}
\end{figure}

Junctions between multiple ponderomotive tubes have been demonstrated in larger, multi-substrate traps \cite{Hensinger06.APL.88.034101,Blakestad09.PRL.102.153002},  but these designs are not as convenient for scaling as the surface-electrode traps, in part because of the difficulty of alignment and assembly. An ideal junction would produce an rf pseudopotential zero at the axis of the pondermotive confining tube along all three legs, merging in the junction center.  Since such an exact pondermotive zero is not possible for a Y junction \cite{Wesenberg09.PRA.79.013416}, we numerically optimized the shape of the rf rails in the junction (see section~\ref{optimization}) to minimize the magnitude of the ponderomotive potential along a continuous path to the center of the junction.  An iterative algorithm began with an initial shape (figure~\ref{Optimization}(a)) to generate the design shown in figure~\ref{Optimization}(c).  Other choices for optimization criteria generate alternate junction geometries that might have benefits over the design that is used in this trap.  For example, as was shown in \cite{Blakestad09.PRL.102.153002}, efficient transport through a junction is possible even if the junction has large deviations from the ideal of an exact pondermotive zero. 

The control electrodes for the junction are narrow near the junction center for increased spatial control and some electrodes are connected on the surface to allow for connections from other parts of the trap to pass under the junction. The final junction design was added to the component library for assembly into the hexagonal ring.

\subsection{Design library}

\begin{figure*}
  \hspace{1in}\includegraphics{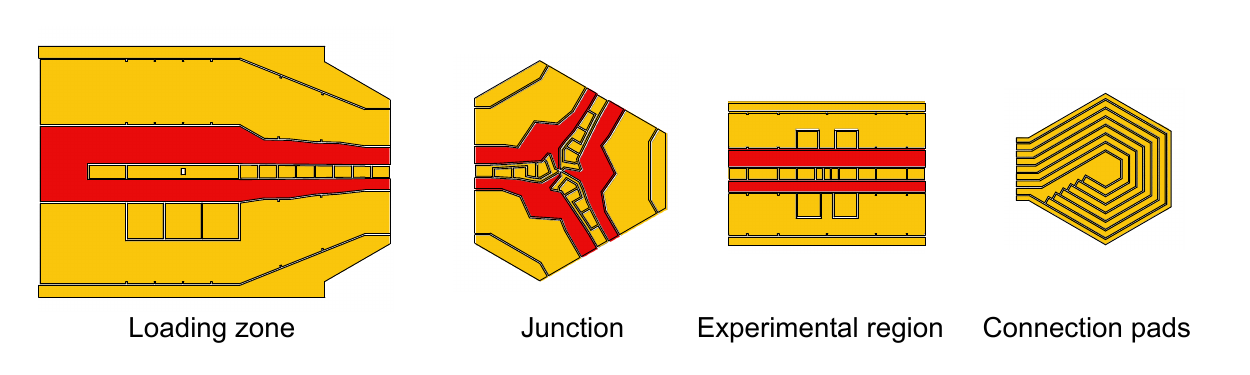}
  \caption{Example components from the design library. The components are assembled to form larger trapping structures such as the design detailed in this paper. Additional components provide standardized electrical connections on the lower conducting layer for the control electrodes.}
  \label{library}
\end{figure*}

The trap geometry shown in figure~\ref{image}(c) was assembled from a library of component designs. Each component has a specific function such as loading or transporting ions. A sample from this library is shown in figure~\ref{library}.  By developing and testing a component library, other trap designs can be assembled from the same library and use precalculated transport waveforms (see section~\ref{waveforms}). This components library includes connections to the control electrodes using the lower conducting layer.  Standardizing the connections in this way greatly simplifies the design process since the connections tend to be very dense around complex components such as the junctions.    
 
\subsection{Electrode geometry optimization}\label{optimization}

The optimization procedure for the junction shape begins with a Y junction that has  straight 
rf electrodes, as shown in figure~\ref{Optimization}(a). The two rf rails in each arm of the 
Y junction have different widths, 40 and 60 $\mu$m, respectively, in 
order to rotate the rf quadrupole axis. This rotation simplifies the control potentials 
needed to allow a single laser beam to Doppler cool a trapped ion \cite{Amini09}. Close to the junction, the minimum value of the 
rf ponderomotive potential is not zero at every point along the center of the
ponderomotive tube, as calculated in the gapless plane approximation \cite{Wesenberg08.PRA.78.063410,House08.PRA.78.033402} but instead forms rf `bumps', as shown in figure~\ref{Optimization}(b).  
The optimization objective was to find appropriate 
electrode shapes that minimized the absolute height of these bumps.  
Eighteen vertices evenly spaced over the inner and outer edges of the 
electrodes, serve as optimization parameters.  The Nelder-Mead 
simplex optimization algorithm \cite{Press02.NumericalRecipes} was then
used to systematically move the vertices around in order to minimize the
height of the bumps while maintaining the threefold rotational symmetry 
of the junction.  Figure~\ref{Optimization}(c) illustrates
the optimized electrode shapes and the corresponding ponderomotive
bumps are shown in figure~\ref{Optimization}(d), achieving a suppression factor of greater than $300$ over 
the initial ponderomotive bump height in figure~\ref{Optimization}(b).

The control electrodes were designed once the rf electrodes shapes were determined.  For adequate control along the ponderomotive tubes in the straight transport sections, the control electrodes were divided into 60 $\mu$m long segments (roughly 1.5 times the ion/surface distance). Near the junction, the control electrodes were designed with narrower electrodes to increase the spatial control of the potentials.  The junction design also allows connections from the inside of the hexagonal ring to the outside to pass under the control electrodes.

\subsection{Transport waveforms}\label{waveforms}

We apply time-dependent potentials (waveforms) to the control electrodes to transport 
the ions along the ponderomotive tube.  The process of
generating these waveforms begins with a calculation to determine the ponderomotive tube center at 
1 $\mu$m intervals along the region the ion is to be transported. 
The goal then is to find the appropriate control electrode 
potentials that trap the ion in a harmonic potential well
centered at each of these points.  The potentials are constrained to not 
displace the ion transversely to the tube, where it will experience rf 
micromotion \cite{Amini09}.  
Smoothly switching between the potentials 
associated with successive wells will then form a waveform that
transports the ion along the rf tube.  

The process of calculating these potentials is divided into two steps for
each of the 1 $\mu$m spaced locations. In the first step,
we use constrained linear programming optimization to 
ensure that there is no electric field at the desired ion location.
Given $n$ participating control electrodes, let $\mathbf{A}$ 
represent the $3\times n$ matrix containing
in the $j^{\textnormal{th}}$ column the contributions to the electric 
field component in the $x$, $y$, $z$ directions, respectively, at the
center of the well if the $j^{\textnormal{th}}$ electrode is held at $1$ 
V and all others at $0$ V.
The  null-space of $\mathbf{A}$ then gives a set of $n-3$ vectors 
$\mathbf{v}$, each satisfying the condition
$\mathbf{Av}=0$, i.e. the entries in each vector $\mathbf{v}$ gives a 
combination of potentials that may be applied to
 the $n$ electrodes without producing an electric field at the well 
center. The desired potential well may be
constructed from a linear combination of these vectors while maintaining 
the zero field condition.  We also constrain the potentials to be 
in the range $\pm$ 5 V, which is within the range our control electronics
can generate.

In the second step, the potential well of the desired trapping frequency 
is now generated from a linear combination of the above solutions.
We use a Nelder-Mead simplex 
optimization algorithm \cite{Press02.NumericalRecipes}, which varies the contribution
of each nullspace vector in such a linear combination.  In each 
iteration the potential due to the current linear combination
of nullspace vectors is calculated at ten points along the 
tube, extending  over $\pm$ 50 $\mu$m from the well center.  
The algorithm minimizes the root-mean-square 
difference between the calculated potential and a
target harmonic potential. This method of fitting to
the $\pm$ 50 $\mu$m interval ensures that the well
has sufficient depth along the rf tube.

Ideally the potentials calculated by the above method would vary smoothly 
as the ion is transported from well-center to well-center, but we find that the
algorithm intermittently produces sharp, unwanted jumps in the 
potentials.  To remove these jumps, the potentials are
post-smoothed by replacing the potentials associated with alternating 
well-centers by the average of the potentials
associated with the two nearest neighboring wells.  After ten levels of 
smoothing the zero-field condition at the well-center is only weakly 
violated. We estimate this violation to be no greater than that caused 
by uncertainties in the trap fabrication, which limits the precision 
to which we can specify the location of the tube center to within 1 $\mu$m.

\section{Results}

To test the basic features of the trap, $^{24}$Mg$^+$ ions were loaded by photoionizing (285 nm) a neutral flux of $^{24}$Mg created in a resistively-heated oven. To prevent contamination and possible shorting of the trap electrodes by the neutral flux,  the oven was located behind the chip (as viewed in Figs. 1(b) and (c)).  Some of the flux then enters the loading regions through two 12 $\mu$m by 20 $\mu$m slots (see figure~\ref{ChipSection}(b)).  From 1 to 10 ions were typically loaded, depending on the loading duration, neutral flux, and photoionization intensity. The ponderomotive potential is significantly perturbed in the vicinity of the loading slots, primarily because the dielectric quartz distorts the rf electric field. After loading, the ions were located 20 $\mu$m to the side of the slot (along the axis) due to this perturbation.   

The trap was enclosed in a vacuum chamber that had several glass viewports for laser access and for imaging the ions. The viewport used for the imaging was 6.5 cm in diameter and located 2.5 cm above the trap surface. We suspect that photoelectrons emitted from the trap surface by the photoionization beam and the Mg$^+$ Doppler cooling beams (280 nm) were charging the glass in the viewport. This apparent charging prevented the trap from holding ions without applying strong compensation fields using the control electrodes. The time scale for the charging was minutes while discharging required several hours. To mitigate this problem, we installed a grounded gold screen (42 $\mu$m spacing with 83 \% light transmission) 5 mm above the trap, parallel to the surface.   At this distance, the screen does not significantly affect the trap potentials but it greatly reduced the apparent charging effects \cite{Pearson06.PRA.73.032307}. 

\subsection{Transport}

\begin{figure}
  \hspace{1in}\includegraphics{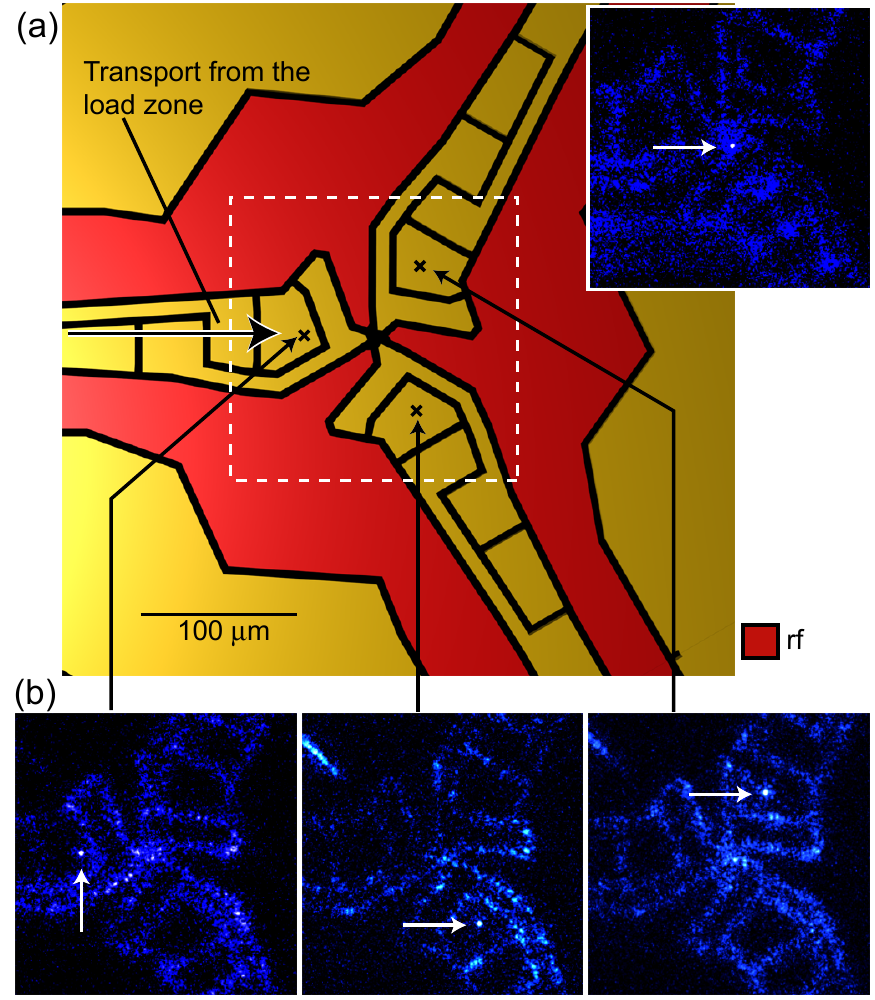}
  \caption{(a) Diagram of a Y junction with an inset showing an ion trapped in the junction center. (b) Transport through the Y junction.  Each photo (false color) shows an ion in one of three locations around the junction. All ions were transported to the junction from a load zone attached to the junction's left leg. To illustrate the electrodes locations in the photos, a laser beam was directed at the quartz substrate so that the scattered light illuminates the gaps between the electrodes.}
  \label{ytransport}
\end{figure}

Ions were successfully loaded in both loading regions of the trap and transported to and through the first adjacent Y junction, as shown in figure~\ref{ytransport}.  We transported individual or groups of ions 850 $\mu$m from the loading zone to the leg of the junction in 2 ms (figure~\ref{ytransport}(a)).  Potentials applied to the control electrodes then transfer the ion to either of the other two legs. Figure~\ref{ytransport}(b) shows a trapped ion in three locations around the leftmost junction of figure~\ref{image}.  We were able to transport ions between any pair of the junction's three legs and in any order.    We also transported ions to the center of the junction  directly from the loading zone, as shown in figure~\ref{ytransport}(a). Transport between the legs in this initial test was ballistic and required that a laser cooling beam overlap the final location of the transport in each leg. We believe that discrepancies between the calculated and observed potentials, as determined from the observed ion position, near the junction center due to time-varying stray fields prevented us from using adiabatic transport \cite{Blakestad09.PRL.102.153002} between the junction legs.

\subsection{Heating rate}

One characteristic parameter of an ion trap is the in-situ heating rate of the ions due to fluctuating electric fields at the ion location.  Typically, miniature rf Paul traps have an observed heating rate orders of magnitude greater than expected from simple Johnson noise on the control electrodes \cite{Deslauriers04.PRA.70.043408,Epstein07.PRA.76.033411,Amini09}.  The source of these noisy electric fields is still not understood, but they can be approximately characterized according to ion-electrode distance. The heating rate for a single ion at the location shown in figure~\ref{image}(c) was observed to be $\dot{n} = 87(11)\times10^3$ phonon/s at a 3.5 MHz axial frequency and a 38 $\mu$m ion-to-surface distance, using the recooling method described in \cite{Epstein07.PRA.76.033411}. The corresponding electric field noise spectral density seen by the ion is $S_E(\omega) \approx \dot{n}(4m\hbar\omega/e^2) \approx 1.3(2)\times10^{-9}$ V$^2$m$^{-2}$Hz$^{-1}$ \cite{Turchette00.PRA.61.063418, Epstein07.PRA.76.033411}, where $m$ and $e$ are the ion mass and charge, respectively. This places this trap in the average range of heating rates for its size scale. We also characterized a separate linear surface-electrode trap made with 1 $\mu$m thick electroplated gold.  The linear trap was designed with a subset of the component library used for the trap of figure~\ref{image}. The heating rate in this trap gave $S_E(\omega) < 6\times10^{-11}$ V$^2$m$^{-2}$Hz$^{-1}$ as determined by the axial heating of a $^{24}$Mg$^+$ ion at 4.5 MHz and 40 $\mu$m  using the recooling method.

\section{Conclusions}

Remaining issues that need to be investigated with this type of trap include the sources of stray electric fields and apparent discrepancies between the calculated potentials and the observed ion positions.  The design components are not limited to the gold-on-quartz construction and could be realized with other choices of materials and fabrication processes. Different strategies for electrode geometry optimization should also be explored. We expect that as the component library is refined and expanded, the design and construction of ion traps for future quantum information processing experiments will become considerably faster and more reliable.  

\section*{Acknowledgments}
\addcontentsline{toc}{section}{Acknowledgments} 

We thank M.G. Blain, J.J. Hudgens, A. Pimentel, and J. Gallegos of Sandia National Laboratories for the 
focused ion beam milling of the loading slots and wirebonding. We acknowledge funding by IARPA, ONR, and the NIST quantum information program.
H.U. acknowledges support from the South African Council for Scientific and Industrial Research (CSIR). 
This paper is a contribution of NIST and is not subject to U.S. copyright.

\section*{References}
\addcontentsline{toc}{section}{References} 

\bibliographystyle{nature}
\bibliography{RaceTrackTrap}

\end{document}